\documentclass[preprint]{aastex6}
\usepackage{graphicx}
\usepackage[figuresright]{rotating}
\usepackage{epsfig}
\usepackage{apjfonts}
\shorttitle{The Thermonuclear Runaway and the Classical Nova Outburst}
\shortauthors{Starrfield, Iliadis, Hix}
\begin{document}

\title{The Thermonuclear Runaway and the Classical Nova Outburst}

\author{S. Starrfield\altaffilmark{1}, C. Iliadis\altaffilmark{2},
W. R. Hix\altaffilmark{3}}

\altaffiltext{1}{School of Earth and Space Exploration, Arizona
State University, Tempe, AZ 85287-1404: starrfield@asu.edu}
 \altaffiltext{2}{Department of Physics and
Astronomy, University of North Carolina, Chapel Hill,
NC27599-3255: iliadis@unc.edu} 
\altaffiltext{3}{Physics Division, Oak Ridge
National Laboratory, Oak Ridge, TN 37831-6354 \& Department of
Physics and Astronomy, University of Tennessee, Knoxville, TN
37996-1200: raph@ornl.gov} 

%\pagenumbering{roman}
 %\maketitle
%\tableofcontents
%\cleardoublepage
%\pagenumbering{arabic}
\def\etal{{et al.\ }}

\begin{abstract}
Nova explosions occur on the white dwarf component of a
Cataclysmic Variable binary stellar system that is accreting
matter lost by its companion.  When sufficient material has been
accreted by the white dwarf, a thermonuclear runaway occurs and
ejects material in what is observed as a Classical Nova explosion.
We describe both the recent advances in our
understanding of the progress of the  outburst and outline some
of the puzzles that are still outstanding.  
We report on the effects of improving both the nuclear reaction rate
library and including a modern nuclear reaction network in our
one-dimensional, fully implicit, hydrodynamic computer code.  In addition, there has been
progress in observational studies of Supernovae Ia with implications
about the progenitors and we discuss that in this review.
\end{abstract}

\section{Introduction}

 The Classical Nova (CN) outburst is one consequence of the
accretion of hydrogen-rich material onto a white dwarf (WD) in a
close binary system. Over long periods of time, the material
being accreted from the secondary star 
forms a layer of nuclear fuel on the WD.  The bottom
of this layer is gradually compressed by the
surface gravity of the WD and ultimately it
becomes electron degenerate.  The degeneracy of the material acts to
prevent the material from expanding even as the temperatures increase from both
compression and nuclear fusion.  Once the temperature at the bottom of the
accreted layer reaches
the Fermi temperature ($\sim 7 \times 10^7$K) the material can expand but by
this time the temperature is increasing so rapidly that a
thermonuclear runaway (TNR) results.  As a result the
temperatures in the nuclear burning region will exceed $10^8$ K
for the lowest mass WDs ($\sim 0.6$M$_\odot$) and possibly reach
$\sim 4 \times 10^8$K  for WDs near the Chandrasekhar Limit.  Further, a major
fraction of the nuclei in the envelope capable of capturing a
proton (CNONeMg...) are transformed into $\beta^+$-unstable
nuclei ($^{13}$N, $^{14}$O, $^{15}$O, $^{17}F$). The $\beta^+$-decay
time scales  limit nuclear energy generation on the dynamical
timescale of the TNR (a few hundred seconds) and their decays at late times produce
extremely non-solar CNO isotopic abundance ratios in the ejected gases. 

Observations of the outburst show that a CN explosively ejects
metal enriched gas and this material is a source of
heavy elements for the Interstellar Medium (ISM).  In some CNe
grains form in the ejecta once the expanding gas has cooled to
temperatures of $\sim$ 1500K some 50 to 100 days into the 
outburst \citep {starrfield_1997_aa, gehrz_1998_aa, jose_2004_aa}. The observed
amount of metal enrichment in the ejected gases demands that mixing of the accreted
material with core material occur at some time during the
evolution of the outburst. The velocities measured for CN
ejecta exceed, in many cases, $10^3$ km s$^{-1}$ so that this
material is rapidly mixed into the diffuse interstellar gas and then
incorporated into molecular clouds before being formed into young
stars and planetary systems during star formation. Therefore, CNe
contribute to Galactic chemical evolution. They
are predicted to be the major source of $^{13}$C, $^{15}$N and
$^{17}$O in the Galaxy and may contribute to the abundances of other
isotopes in the intermediate mass range \citep{gehrz_1998_aa}.

Infrared observations have confirmed the formation of carbon, SiC,
hydrocarbons, and oxygen-rich silicate grains in CN ejecta,
suggesting that some fraction of  the pre-solar grains identified
in meteoritic material  \citep{zinner_1998_aa, amari_2001_aa} 
and anomalous interplanetary grains \citep{pepin_2011_aa} may come
from novae \citep {starrfield_1997_aa, gehrz_1998_aa, jose_2004_aa, pepin_2011_aa}.  
Observations imply that the mean mass
ejected during a CN outburst is $\sim 2\times 10^{-4}$ M$_\odot$
\citep{gehrz_1998_aa}.  Using the observed CN rate of 35$\pm$11 per year in our
Galaxy \citep{shafter_1997_aa, shafter_2002_aa}\footnote{Shafter (2015, priv. comm.) now 
believes that this number is a lower limit and $\sim$ 50 is more reasonable.}, it follows that they introduce $\sim
7\times 10^{-3}$ M$_\odot$ yr$^{-1}$ of processed matter into the
ISM.  However, this value may be a lower
limit \citep{saizar_1994_aa, gehrz_1998_aa}.

In the next section (2), we describe how the TNR is initiated and follow that with sections on: (3) the initial
conditions, (4) the effects of new reaction rates, (5)
multidimensional studies of the TNR, (6) nucleosynthesis and  
the ejecta mass discrepancy, and (7) the proposed
relationship of CNe to the progenitors of Supernovae of Type Ia.
We end with a Summary and Discussion.

\section{Initiating The Thermonuclear Runaway}

Hydrodynamic studies have shown that the consequences of accretion
from the secondary is a growing layer of hydrogen-rich gas on the
WD. When both the initial WD luminosity  and the rate of mass accretion
onto the WD are sufficiently low ($L \le10^{-2}$L$_\odot$ and \.M$ \le 10^{-9}$M$_\odot$~yr$^{-1}$)
a layer of unburned hydrogen-rich
gas ($\sim 10^{-4}$ M$_{\odot}$ to $\sim 10^{-6}$ M$_{\odot}$, 
a decreasing function of increasing WD mass) can accumulate on the WD surface.  
Both compressional heating and the energy released by nuclear
fusion (once the temperatures at the bottom of the accreted layers have reached a few million degrees)
heat the accreted material.  Since the deepest layers of
the accreted material have become both hot and electron
degenerate, the temperatures will rise with little or no expansion of
these layers.  At a temperature of $\sim 7 \times 10^7$K the degeneracy
becomes unimportant and the layers can begin to expand.  However, by this
time the temperature is increasing so rapidly a TNR occurs and it takes only
a few hundred seconds or less for the temperatures to reach a peak value which
depends on the mass of the WD.  

For the physical
conditions of temperature and density that occur in this
environment, nuclear processing proceeds by hydrogen burning,
first from the proton-proton chain [including the {\it pep}
reaction: $p + e^{-} +p \rightarrow d + \nu$ \citep{schatzman_1958_aa, bahcall_1969_aa}
which plays a significant role \citep{starrfieldpep09}]
and, subsequently, via the CNO cycles. If there are heavier nuclei
present in the nuclear burning shell, then they will contribute
significantly to the nucleosynthesis.  The range of peak
temperatures typically sampled in CN outbursts ($10^8$K to $\sim 4 \times 10^8$K
depending on WD mass) gives
rise to significant energy production. 

The proton-proton chain is important during the main accretion
phase of the outburst when the amount of mass accreted
prior to the TNR is determined.  It is the CNO-cycle reactions, however, and, ultimately,
the hot CNO cycles that power the final stages of the TNR and
the evolution to the peak of the explosion.  Energy production and
nucleosynthesis associated with the CNO cycles
impose important constraints on the energetics of the
runaway. In particular, the rate of nuclear energy generation at
high temperatures (T $>$10$^{8}$ K) is limited by the timescales
of the slower, temperature insensitive, $\beta^+$-decays,
particularly $^{13}$N ($\tau_{1/2}$ = 598 s), $^{14}$O
($\tau_{1/2}$ = 71 s), $^{15}$O ($\tau_{1/2}$ = 122 s), 
and $^{17}$F ($\tau_{1/2}$ = 64 s).  The
behavior of the $\beta^+$-decaying nuclei holds important
implications for the nature and consequences of CN outbursts.  For
example, significant enrichment of CNO nuclei in the nuclear
burning regime is required to insure high levels of energy release
on a hydrodynamic timescale (seconds for WDs) and thus produce a
violent outburst \citep{starrfield_1989_aa, starrfield_1998_aa, jose_1998_aa, yaron_2005_aa,
starrfield_2008_CN, starrfieldpep09}.

The large abundances of these positron emitters, at the peak of
the outburst, have important consequences for the
evolution:
  
\begin{itemize}

\item When temperatures in the nuclear burning region significantly
exceed $10^8$ K, proton captures transform CNO nuclei to the positron 
emitters $^{13}$N, $^{14}$O, $^{15}$O, and $^{17}$F.

\item Since the energy production in the CNO cycle comes
from proton captures, followed by $\beta^{+}$-decays, the rate
of nuclear energy generation, at temperatures exceeding 10$^{8}$ K,
depends only on the half-lives of the positron emitters and
the numbers of CNONeMg nuclei initially present in the envelope.

\item  At temperatures exceeding $10^8$ K, 
the convective region ranges from the bottom of the nuclear burning
region up to nearly the surface of the
accreted envelope bringing unburned
CNONeMg nuclei into the nuclear burning region when the temperature is
rising extremely rapidly.  This process keeps the nuclear reactions
operating far from equilibrium. 

\item Since the convective turn-over
time scale can range from  10 to $10^{2}$ s near the peak of the TNR, a
significant fraction of the radioactive nuclei reach the surface of the WD. 
Their decays at the surface yields a nuclear energy generation rate of
10$^{13}$ to 10$^{15}$ erg
g$^{-1}$ s$^{-1}$ (depending upon the enrichment).  

\item Their half-lives are longer than the hydrodynamic expansion time of the
outer layers and thus the radioactive nuclei decay when the temperatures in
the envelope have declined to values that are too low for any
further proton captures to occur, yielding isotopic ratios in the
ejected material that are distinctly different from the ratios
predicted from the equilibrium operation of the CNO cycles. 

\item The
decays of the radioactive nuclei provide an intense heat source throughout
the envelope that flattens the temperature gradient and ultimately
shuts off convection.

\item Finally, the energy release from the
$\beta^+$-decays throughout the envelope helps eject the material from the WD.

\end{itemize}

Hydrodynamic studies of CN explosions show that, if core material
is mixed into the accreted material, then sufficient energy
is produced during the evolution described above, to eject
material with expansion velocities that agree with observed values.
Further, the predicted bolometric light curves for the early
phases are in reasonable agreement with the observations
\citep{starrfield_1989_aa, starrfield_1998_aa, gehrz_1998_aa}
as are the nucleosynthesis predictions \citep{jose_1998_aa, jose_2004_aa, starrfieldpep09}.  
The hydrodynamic studies also show that
at least three of the observational behaviors of the CN outburst
are strongly dependent upon the interaction between
nuclear fusion and convection that occurs during the final
minutes of the TNR. These are: (1) the early evolution of the
observed light curves of CNe on which their use as ``standard
candles'' is based. (2) The observed peak luminosity of fast novae
which is typically super-Eddington (in some cases for as long as
two weeks; see, {\it e.g.}, Schwarz et al. 2001; Quataert et al. 2015)\nocite{quataert_2015_aa}. 
(3) The composition of matter
ejected in a CN outburst which depends on the amount and composition of the
material dredged up from the underlying CO or ONe WD core.  We emphasize that the
existence of this mixing is demanded by observations of CNe ejecta
\citep{gehrz_1998_aa, jose_2004_aa, downen_2012_aa}.  

Predicting the ejecta composition is also critical to
questions concerning the possibility of observing nuclear decay $\gamma$-rays
(from $^7$Be and/or $^{22}$Na) from nearby CNe  \citep{hernanz_2008_cn},
and the contributions of CNe both to Galactic chemical evolution
and to the isotopic anomalies observed in some pre-solar grains
\citep{amari_2001_aa, jose_2004_aa, jose_2007_ab}, and Anomalous Interplanetary Particles \citep{pepin_2011_aa}. 
Moreover, the amount of core matter in the
ejecta implies that the WD in a CN system is losing mass as a
result of continued outbursts, and thus it has been argued that a CN system cannot be a SN Ia progenitor
\citep{macdonald_1984_aa, starrfield_2000_aa}.  This, however, may
not be the case for typical Cataclysmic Variables in which the accreted material
does not appear to be mixing with core material \citep{starrfield_2014_aa}

As already mentioned, the $\beta^{+}$-decay heating of the
outermost regions of the nova envelope reduces the temperature
gradient and, in turn, reduces convection in the surface layers
around the time of peak temperature in the TNR. The growth of
convection from the burning region to the WD surface and its
subsequent retreat in mass, as the envelope relaxes from the peak
of the TNR on a thermal timescale, implies that considerable 
variations in the elemental and isotopic
abundances should exist in the ejected gases. Observations that provide either
abundance gradients or isotopic abundances in CN ejecta can critically constrain our
knowledge of both the amount of mixing and the history of convection during the TNR.

\section{The Physical Processes that Affect the Amount of Accreted Material}

The history of the TNR hypothesis for the CN
outburst was described in \citet{starrfield_1989_aa}, and will not be repeated
here. One of the important developments since that review 
was published were the various calculations of the amount of hydrogen-rich material
required to trigger the TNR.  In the 1980's, there
were both analytic \citep{fujimoto_1982_ab, fujimoto_1982_aa} and semi-analytic
\citep{macdonald_1983_aa} calculations to determine the amount of material. Since
that time, there have been a number of studies of accretion onto
WDs using Lagrangian hydrodynamic computer codes
to follow the evolution of the material as it is
accreted onto the WD \citep[][and references therein]{starrfield_1998_aa, jose_1998_aa,
starrfield_2000_aa, yaron_2005_aa}. These calculations show that the amount of
material accreted onto the WD depends on the WD mass, the WD
luminosity, the composition of the accreted matter, and the
rate of mass accretion.   

Theoretical studies have also shown that the
characteristics of the outburst also depend on the initial luminosity 
and thermal structure of the WD
\citep{townsley_2004_aa} and a higher initial luminosity results in less
mass being accreted.  Repeated outbursts on a WD can also change the
thermal structure affecting the amount of accreted material and, therefore,
the evolutionary history of the WD is a fifth parameter that is important in
understanding the CN outburst.  If
mixing of accreted material with core material occurs during the
accretion phase, then the opacity in the nuclear burning region
increases and traps more heat in this region than if no mixing
has occurred.  As a result, the temperature in the nuclear burning
region increases rapidly, reduces the time to TNR and
thereby the total amount of accreted and ejected material
\citep{starrfield_1998_aa}.  We note, however, that recent multi-dimensional
studies imply that the mixing does not occur until close in time to the peak
of the TNR \citep{jose_2014_aa}.  Since the amount of accreted material
directly affects the characteristics of the outburst, a single valued ``maximum
magnitude rate of decline'' (MMRD) relationship does not exist as
is shown by observations \citep{kasliwal_2011_aa}.

Given that the evolution begins with a WD, that has
a surface layer rich in helium remaining from previous
outbursts \citep{shara_1989_aa, krautter_1996_aa,starrfield_1998_aa} most of the time
is spent, and most of the mass is accreted, during the phase when
the principle energy production mechanism is the proton-proton
chain \citep{starrfield_1998_aa, starrfield_2000_aa, starrfieldpep09}. 
During this evolutionary phase, there is a
competition between the energy production, which has an X$^2$T$^{4-6}$
dependence (X is the hydrogen mass fraction), degenerate electron conduction into the interior, and
radiative diffusion to the surface. Since the thickness of the
surface layers is small and convection is not yet important, most
of the energy produced at the bottom of the accreted layers is
transported to the surface and radiated, while only a small
fraction is transported into the interior.  Therefore, the
temperature in the nuclear burning region increases 
slowly as mass is accreted.  However, the inclusion of the  $pep$ reaction ($p + e^{-} +p \rightarrow d +
\nu$: \citet{schatzman_1958_aa, bahcall_1969_aa}) increases the rate of energy generation
at a given temperature and density so that less mass is accreted and peak temperatures are
lower.  As reported in
 \citet{starrfieldpep09}, including the $pep$ reaction on a 1.25M$_\odot$ WD reduces the
 accreted mass by $\sim$30\% (T$_{peak}$ is reduced by $\sim$ 8\%) and $\sim$10\% on a 1.35M$_\odot$ WD
 (T$_{peak}$ is reduced by $\sim$ 11\%).

The amount of mass accreted during proton-proton chain burning also
depends on the metallicity of the material.  Increasing the metallicity of the
accreting material results in an increase in the opacity. The
increased opacity results in more heat, produced by
compression and nuclear burning in the deeper layers of the
accreted material, being trapped in the region where it is 
produced so that the temperature increases faster per unit accreted
mass than in a simulation with a lower metallicity and opacity \citep{starrfield_1998_aa, jose_2007_aa}.
In contrast, lowering the metallicity by accreting material
representative of the LMC (one-third Solar metallicity or less),
reduces the opacity and increases the rate of radiative heat
transport out of the nuclear burning layers.  As a result, the
temperature increases more slowly than for higher metallicity
material and more material is accreted.  A more massive accreted
layer implies a higher density at the bottom and a more violent
explosion \citep{starrfield_1998_ae, starrfield_1999_aa, yaron_2005_aa, jose_2007_aa}. This result is in agreement
with the observations of CNe in the LMC \citep{dellavalle_1992_aa, dellavalle_1994_aa, schwarz_2001_aa}.

If, the accreted material mixes with core material
during the proton-proton chain burning phase, either by shear mixing \citep{kuttersparks_1987_aa, 
sparks_1987_aa, rosner_2001_aa, alexakis_2004_aa} 
or by elemental diffusion \citep{prialnik_1984_aa, kovetz_1985_aa, yaron_2005_aa}
then the heavy nuclei will be enriched in the accreted
layers and, in turn, the opacity in the nuclear burning
layers will increase.  This enrichment will reduce the amount of material
accreted before the onset of the TNR and, thereby, the amount of
material ejected during the outburst. Given that the theoretical
predictions of the amount of material ejected during the outburst
are lower then observed, increasing the amount of metals in the
accreted layers by early mixing exacerbates this disagreement
arguing for mixing to occur late in the accretion phase.  
In fact, the multi-dimensional studies of mixing at a late stage in the evolution to the
TNR when convection is already important, 
\citep[see][and references therein]{jose_2014_aa} are sufficient to produce
the amount of core material in the ejecta and, because the mixing occurs late
in the evolution, the amount of  
metals in the nuclear burning region 
have no affect on the amount of accreted material.  In this case the
studies of the accretion of Solar material are relevant to the 
amount of material accreted by the WDs of various masses \citet{starrfield_2014_aa}.

There is an interesting corollary to this discussion.  As the
opacities have been improved (more levels, better line profiles,
more elements included, better equations-of-state) by the various
groups working in this area 
\citep[see, for example:][]{rogers_1994_aa, iglesias_1996_aa, rogers_2002_aa},
they have also increased for a given temperature and
density irrespective of the metallicity.  We have found that the amount of accreted material
has decreased with the inclusion of modern opacities.
Therefore, even without mixing core material with accreted
material or changing the metallicity of the accreting material,
recent simulations have increased the discrepancy between
theory and observation with respect to the amount of ejected
mass \citep{starrfield_2000_aa, starrfieldpep09}.

In order to better study this effect, we updated and improved, NOVA, our 1D,
hydrodynamic, evolution code by including the latest OPAL
opacities \citep{rogers_1994_aa, iglesias_1996_aa, rogers_2002_aa}.  
We calculated new evolutionary
sequences for 1.25M$_\odot$ WDs, in an attempt to simulate the
outburst of V1974 Cyg \citep{starrfield_1998_aa}.  The revised
opacities had profound effects on the simulations. Because the
modern opacities were larger than those we had been using [the
\citet{iben_1982_aa} fit to the \citet{cox_1970_aa, cox_1970_ab} 
and the \citet{cox_1976_aa} opacities], we found that our new simulations 
ejected a factor of ten less mass
than was inferred from observations of the outburst of V1974 Cyg
\citep{starrfield_1998_aa, vanlandingham_2005_aa}. This discrepancy was also found in
a study of accretion onto ONe WDs \citep{jose_1997_aa, jose_1998_aa}.
In \citet{starrfield_1998_aa}, we proposed a possible solution to this problem.  As already
mentioned, the WD spends a major fraction of time during the accretion phase
generating energy from the proton-proton chains for 
which $\epsilon_{nuc} \approx  X^2 T^{4-6}$.  Any
change in the physical conditions that lengthens the time spent in
this phase will increase the accreted mass.  Mixing of the accreting hydrogen-rich material
into a residual helium enriched shell \citep[the remnant of previous
outbursts:][]{krautter_1996_aa} would reduce both
the hydrogen mass fraction and the opacity slowing the rise in temperature 
and allowing more mass to be accreted.

\citet{prialnik_1982_aa} were the first to show a
strong effect of the rate of mass accretion on the ignition mass.
They reported that increasing the rate of mass accretion increased
compressional heating and, thereby, caused the temperature in the
accreted layers to rise more rapidly (per unit accreted
mass) than for lower mass accretion rates.  We have found that mass
accretion rates of $\sim 10^{-9}$M$_\odot$ yr$^{-1}$  
\citep{townsley_2004_aa, starrfield_2008_CN}
result in smaller amounts of
material being accreted compared to simulations where the
rate of mass accretion has been reduced by a factor of 10-100.
We also find that increasing the mass accretion rate
above 10$^{-9}$M$_\odot$ yr$^{-1}$, on {\it low} luminosity (L $ \le 10^{-2}$ L$_\odot$) and lower mass
WDs (M$_*$ $ \le 1$ M$_\odot$), causes weak flashes \citep{starrfield_2012_basi, starrfield_2014_aa,
hillman_2015_aa}. 
In this case, the large amount
of heat released by compression keeps the degeneracy low and the
TNR only reaches temperatures of $\sim 10^8$ K.

All other parameters held constant, the internal temperature
(or the observed luminosity) of the underlying WD also affects the amount
of mass accreted prior to the TNR, such that as the
luminosity of the WD declines, the amount of accreted material
increases. There are two reasons that the luminosity is important.
First, in addition to compressional heating from accretion, the
heat flowing from the interior of the WD also heats the region where 
nuclear burning is initiated.
As the WD evolves and cools, this heat source becomes less
important for the accreted layers. In addition, a cooler interior
implies cooler surface layers so that nuclear
reactions begin later in the evolution of the TNR. Once the WD has
undergone a series of outbursts, then its luminosity is determined
by the average rate of mass accretion \citep{townsley_2004_aa}.

The discussion up to this point is most relevant for {\it
low} luminosity WDs.  If the luminosity of the WD is higher,
because either the WD is less evolved or it has not yet reached
quiescence after a CN explosion, then the temperatures in the
surface layers are sufficiently high for nuclear burning to occur
in the accreting material shortly after it arrives on the surface.  
The early nuclear burning drives the WD to an earlier TNR with a 
smaller amount of accreted mass and a less violent outburst.  
More importantly, at some mass accretion rate independent of the
WD luminosity,  the infalling material
is predicted to burn at the rate it is accreted and no TNR results.
\citet{paczynski_1978_aa, sion_1979_aa, fujimoto_1982_ab, fujimoto_1982_aa}
and \citet{iben_1982_aa} introduced the idea of {\em Steady
Burning} which is accretion at a high rate onto WDs. The Steady
Burning mass accretion rate depends on WD mass but typically is a 
few times $10^{-7}$ M$_\odot $ yr$^{-1}$.

Steady burning is important in trying to understand the properties of
the Super Soft X-ray Binaries (SSS).
The SSS were discovered by the Einstein satellite
\citep[two members are CAL 83 and CAL 87:][]{long_1981_aa} but they were
not identified as a stellar class until the ROSAT survey of the
LMC  \citep{trumper_1991_aa}.  SSS are luminous, L$\sim 10^{37}$ erg
s$^{-1}$, with surface temperatures ranging from 30 to 50 eV or
higher \citep{vandenheuvel_1992_aa, kahabka_1997_aa}.  
Optical studies show that they are close binaries
containing a WD \citep{cowley_1998_aa}. 
\citet{vandenheuvel_1992_aa} proposed that
steady burning of the hydrogen-rich material accreted from the secondary
was occurring on the surface of the WD component of the SSS
binary.  As a result, no TNR would occur, no mass would be
ejected, and the mass of the WD could grow to the Chandrasekhar
Limit.   \citet{starrfield_2004_aa} tested their prediction with a series
of evolutionary sequences accreting at high rates but found 
that steady burning occurred only for {\it hot, luminous} WDs.  

\citet{nomoto_2007_aa}, then investigated this problem but used a 
static method devised by
\citet{sienkiewicz_1980_aa} that  is not suitable for evolutionary
studies. \citet{shen_2007_aa} used steady state envelopes 
and neither set of authors could verify the results of \citet{starrfield_2004_aa}.
Further studies, however, suggested that \citet{starrfield_2004_aa} used
too large zone masses for the outermost layers of their calculations
and would have obtained TNRs for smaller zone masses.  This
study is now being redone (Starrfield et al. 2016 in preparation) because, as yet,
there are no evolutionary results that reproduce the observed behavior of
the SSS.
More recently, hydrodynamic simulations of accretion of solar material onto WDs also show
that steady burning, as proposed by \citet{nomoto_2007_aa} and \citet{shen_2007_aa}, does not occur
\citep[][and references therein]{starrfield_2014_aa}.  In fact, as reported by
\citet{newsham_2014_aa} and \citet{ starrfield_2014_aa}, the accretion of 
solar material results in TNRs with only small amounts of mass ejected (see below)
so that the WD is growing in mass for a large range of
WD masses and mass accretion rates. 

Another parameter which affects the amount of material that is
accreted prior to the TNR is the mass of the WD. The amount of material accreted, all other
parameters held constant, is inversely proportional to the mass of
the WD \citep[see, for example][and references therein]{macdonald_1985_aa}. Specific values of the amount of accreted material as a
function of WD mass are given in \citet{starrfield_1989_aa}.
The ignition mass can be estimated from
\begin{equation} P_{\rm crit} = {{\rm G M}_{\rm WD} {\rm M}_{\rm
ign} \over 4 \pi {{\rm R}_{\rm WD}^4}}
\end{equation}

\noindent 
P$_{\rm crit}$ is assumed to be $\sim 10^{20}$ dyne cm$^{-2}$ and
a mass-radius relation for WDs gives the ignition mass, M$_{\rm
ign}$.  Equation 1 is obtained by realizing that a critical
pressure must be achieved at the bottom of the accreted layers
before a TNR can occur \citep{fujimoto_1982_ab, fujimoto_1982_aa, gehrz_1998_aa}. 
However, the actual value of the critical pressure is also a
function of the WD composition and rate of accretion \citep{starrfield_1989_aa}.  
If one assumes the above numerical value for the pressure, then the
amount of accreted mass can range from less than
$10^{-5}$ M$_\odot$ for WDs near the Chandrasekhar Limit to values
exceeding $10^{-3}$ M$_\odot$ for 0.5 M$_\odot$ WDs. In addition,
because the surface gravity of a low mass WD is smaller than for 
a massive WD, the bottom of the accreted layers is considerably
less degenerate when the TNR occurs. Therefore, the peak
temperature, for a TNR on a low mass WD, may not even reach
$10^8$ K and little nucleosynthesis will occur.  In contrast, the peak
temperature on a 1.35 M$_\odot$ WD can exceed $4 \times 10^8$ K
(see Figure 1).

\section{The Effects of Nuclear Reaction Rates on the Outburst}

The behavior of the TNR
depends critically both on the nuclear reactions considered and on
the values of the reaction rates used in the simulations.  It is
the operation of the CNO reactions at high temperatures and
densities that imposes severe constraints on the energetics and
nucleosynthesis of the outburst.   
\citet{starrfieldpep09} presented a study of the impact of steadily
improving reaction rate libraries which included the results of 
ongoing efforts by the nuclear experimentalist community over the
past 20 years.  We briefly summarize their results here.

An additional and important part of \citet{starrfieldpep09} was a switch in the
nuclear reaction rate solver used in, NOVA, 
their 1-dimensional hydrodynamic code, from \citet{weiss_1990_aa} to
the more modern nuclear reaction network solver of \citet{hix_1999_ab} 
\citep[see also][]{paretekoon_2003_aa}.  While both networks utilize
reaction rates in the common REACLIB format and perform their
temporal integration using the Backward Euler method introduced by
\citet{arnett_1969_ab}, there are two important differences.
First, \citet{weiss_1990_aa}
implemented a single iteration, semi-implicit Backward
Euler scheme, which had the advantage of a relatively small and
predictable number of matrix solutions, but allowed only heuristic
checks that the chosen time step provided a stable and accurate
solution.  In contrast, \citet{hix_1999_ab} implemented an iterative, fully
implicit scheme, repeating the Backward Euler step until
convergence was achieved.  The iterations provided both a measure
of the stability and accuracy of the solution.

Second, the \citet{hix_1999_ab} network solver
employs automated linking of reactions in the data set to the
species being evolved. This is in contrast to the manual linking
employed by \citet{weiss_1990_aa} and many older reaction networks.
This automated linking helps to avoid implementation mistakes, as
was discovered while performing tests of NOVA in order to
understand the source of differences in the results of the
simulations between the versions of the code,  which used the
same reaction rate library but different nuclear reaction solvers.
We found that the REACLIB dataset
used in prior studies included the $pep$ reaction 
($p + e^{-} +p \rightarrow d +
\nu$: \citet{schatzman_1958_aa, bahcall_1969_aa}), but it was not linked to
either the abundance changes or the energy generation in the
\citet{weiss_1990_aa} network. 
While for solar models the energy generation from the {\it
pep} reaction is unimportant 
\citep[but not the neutrino losses:][]{rolfs_1988_aa}, in the WD envelope the density can reach, or
exceed, values of $10^4$ g cm$^{-3}$, which increases
the rate of energy generation compared to the simulations done without
the {\it pep} reaction included \citep{starrfieldpep09}. The
increased energy generation reduces the amount of accreted
material since the temperature rises faster per gram of accreted
material.  Given a smaller amount of accreted material at the time when the
steep temperature rise of the TNR begins,  the nuclear burning
region is less degenerate and, therefore, the peak temperatures
are lower when compared to models evolved with the nuclear
reaction rate library used in our previous studies. 

\begin{figure}
\includegraphics[angle=90,scale=0.35]{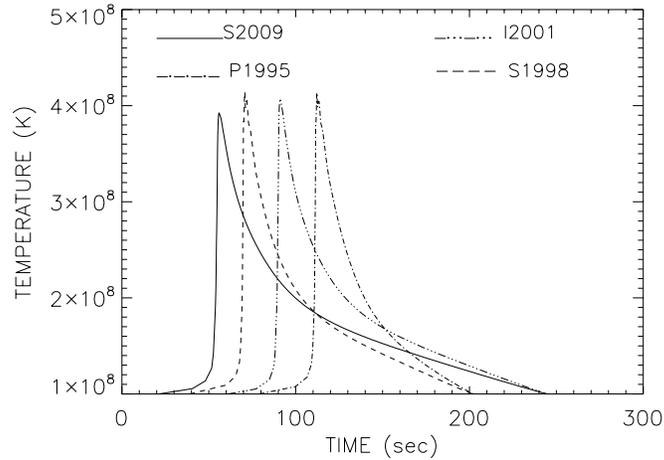}
\centering \caption{The variation with time of the temperature in
the deepest hydrogen-rich zone around the time when peak
temperature occurs.  We have plotted the results for four
different simulations on a 1.35 M$_\odot$ WD. The identification
with a specific library is given on the plot:
P1995 is the library used in \citet{politano_1995_aa}, S1998 is that
used in \citet{starrfield_1998_aa},
I2001 was described in \citet{iliadis_2001_aa} and used in
\citet{starrfield_2001_aa}, and S2009 is the library
used in \citet{starrfieldpep09}.
The temperature declines 
more rapidly for the sequence computed
with the oldest reaction library \citep{politano_1995_aa} because it
exhibited a larger release of nuclear energy throughout the
evolution, which caused the overlying zones to expand more rapidly
and the nuclear burning layers to cool more rapidly.  In contrast,
using the newest library yields the smallest expansion velocities and the
nuclear burning layer cools slowly. }
\end{figure}

Using the \citet{hix_1999_ab} nuclear reaction solver, \citet{starrfieldpep09} used 4 different libraries:

\begin{itemize} 

\item The first library was originally used in \citet{politano_1995_aa} and obtained its
rates from \citet{caughlan_1988_aa} and \citet{thielemann_1986_aa,
thielemann_1988_aa}. The library was provided by F. Thielemann and also used in the
calculations reported in \citet{weiss_1990_aa}.

\item  \citet{starrfield_1998_aa, starrfield_2000_aa} used an updated reaction rate library which
contained new rates calculated, measured, and compiled by
F. Thielemann and M. Wiescher.  A discussion of the improvements over \citet{politano_1995_aa} is 
provided in \citet{starrfield_1998_aa}. 

\item  The third library was described in \citet {iliadis_2001_aa} and was used for the
simulations reported in \citet{starrfield_2001_aa}. 

\item The fourth library was a compilation by Iliadis (2005, priv. comm.) and was current as of August 2005.
It was a major update to the library described in \citet{iliadis_2001_aa}.

\end{itemize}

A detailed discussion of the improvements since \citet{iliadis_2001_aa} appeared 
in \citet{starrfieldpep09}, so we only provide a summary here.  In total, the rates of 11 and 33
proton-induced reactions were adopted from \citet{ang99}  
and \citet{ili01}, respectively.  For 17 proton-induced
reactions, new rates were evaluated based on new
experimental information.  Those included, for example, the
(p,$\gamma$) and (p,$\alpha$) reactions on $^{17}$O, $^{18}$F, and
$^{23}$Na.  A number of rates for $\alpha$--particle induced
reactions, including those for $^{14}$O($\alpha$,p),
$^{18}$Ne($\alpha$,p), and $^{15}$O($\alpha$,$\gamma$), which are
important for following breakout during the hot CNO-cycles, were also
updated. The ground and isomeric state of $^{26}$Al were treated as
separate nuclei \citep{ward_1980_aa} and the communication
between those states through thermal excitations involving
higher--lying excited $^{26}$Al levels was taken explicitly into
account. The required $\gamma$-ray transition probabilities were
adopted from \citet{runkle_2001_aa}.

\citet{starrfieldpep09} evolved seven
different evolutionary sequences for WD masses of 1.25 M$_\odot$ 
and 1.35 M$_\odot$.  Here we only report on the results for the more
massive WD since the differences are more extreme
for this mass.
We assumed an initial WD luminosity of $\sim 4
\times 10^{-3}$ L$_\odot$ and a mass accretion rate of $2 \times 10^{-10}$M$_\odot$yr$^{-1}$(10$^{16}$ g
s$^{-1}$).  This mass accretion rate is 5 times lower than the lowest
rate used in \citet{starrfield_1998_aa} and was chosen to maximize the amount
of accreted matter given the increase in energy generation caused by
including the $pep$ reaction.  
We used the same composition for the accreting material
as used and described in \citet{politano_1995_aa}, \citet{starrfield_1998_aa},
\citet{starrfield_2000_aa}, and \citet{starrfield_2001_aa}: a mixture of half-solar and half-ONeMg
(a mixing fraction of 50\%). 
By using this composition, we assumed that core
material mixed with accreted material from the beginning of
the evolution.  Using this composition also effects the amount of accreted
mass at the peak of the TNR since the opacity is higher when compared to 
simulations that assume no mixing.

The results of the evolutionary sequences for WDs with masses of 
1.35 M$_\odot$ show that because the WD mass is larger and the
radius is smaller, they reach higher densities and higher peak
temperatures than the sequences at lower WD mass \citep{starrfield_1989_aa}.
Figure 1 shows the variation of temperature with time for the
deepest hydrogen-rich zone for four of the 1.35 M$_\odot$ evolutionary
sequences. In this figure, and all other figures in this article, we
only plot the simulations done with the {\it pep} reaction included
(i.e., with the \citet{hix_1999_ab} network).  The specific
evolutionary sequence is identified on the plot and the time
coordinate is arbitrary and chosen to clearly show each curve on the plot.
On this plot and each of the following plots (Figures 1 to 5), the designation refers to
the reaction rate library that was used for the sequence.  They are
P1995 \citep{politano_1995_aa}, S1998 \citep{starrfield_1998_aa},
I2001 \citep{iliadis_2001_aa, starrfield_2001_aa}, and S2009 \citep{starrfieldpep09}.
We see differences between the four simulations since, as we use
a more modern nuclear reaction library, the peak temperature drops
from $4.13 \times 10^8$ K  to $3.92 \times 10^8$ K.

The total nuclear luminosity (in units of L/L$_\odot$)
as a function of time is shown in Figure 2
for 1.35 M$_\odot$. The time coordinate is the same as
in Figure 1.  At 1.35 M$_\odot$ the
maximum luminosity found for the latest library is smaller than found using the
earlier libraries.  The improvements in the libraries are
more important for the heavier isotopes and become even more
important as higher temperatures are reached. The isotopic predictions
for the ejected material in all the sequences are given in \citet{starrfieldpep09}

\begin{figure}
\includegraphics[angle=90,scale=0.33]{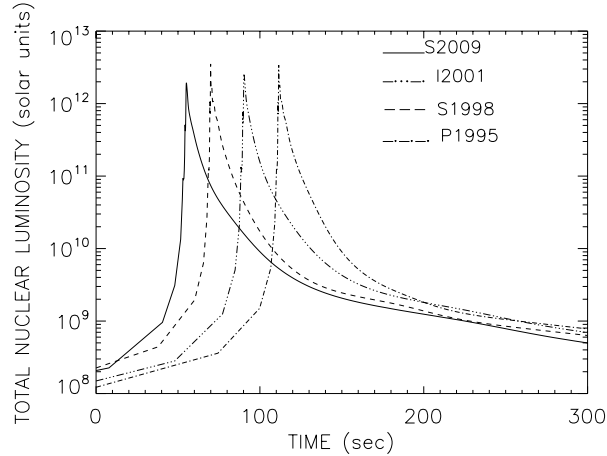}
\centering \caption{The variation with time of the total nuclear
luminosity (L/L$_\odot$) around the
time of peak temperature during the TNR on a 1.35M$_\odot$ WD. We
integrated over all zones taking part in the explosion. The
identification with each library is given on the plot. The time
coordinate is the same as for Figure 1. }
\end{figure}

Figure 3 shows the variation of the effective temperature
with time as the layers begin their expansion.  We have plotted
the results on the same time scale as in Figure 1 and Figure 2 and
the plots show how rapidly the energy and the $\beta^+$-unstable
nuclei reach and heat the surface layers. The large amplitude
oscillations seen in the sequences using the two oldest libraries, but not
seen in the simulations from the two more recent libraries, 
originate from the intense heating at
the surface causing the layers to expand rapidly, cool and
collapse back onto the surface, and then expand again.  The peak
temperatures and luminosities reached in these CN simulations are sufficiently large that
an all-sky X-ray detector would detect them if it were sensitive enough.

The oscillations can be seen more vividly in Figure 4, which shows the
velocity of the outermost layers as a function of time near the
peak of the TNR.   At the beginning of the oscillations, almost no
expansion has occurred and the
``quasi''-period is determined by the free-fall time for the underlying
WD.  The intense heating from the $\beta^+$-decays causes the
luminosity to quickly become super-Eddington and the layers 
begin expanding.  However, they are still deep within the potential
well of the WD and oscillate for a few seconds before reaching and
then exceeding escape velocity at which time the oscillations cease. 
The oscillations are not
present in the latest sequence because surface heating is less
important.  The outburst evolves more gradually and the star has already begun
expanding when the $\beta^+$-unstable nuclei reach the
surface.   As a result, the oscillations occur but are of much smaller amplitude.

\begin{figure}
\includegraphics[angle=90,scale=0.33]{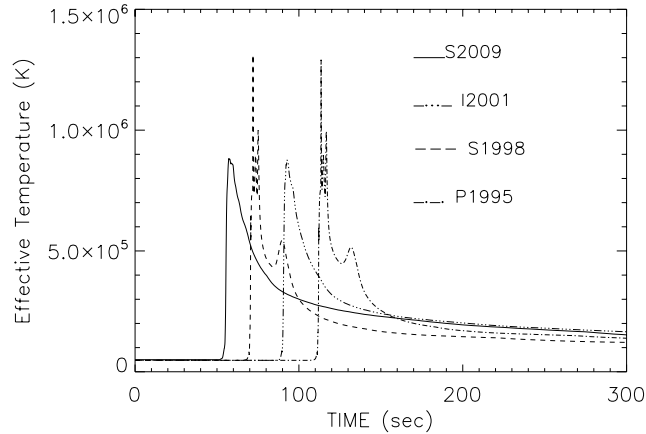}
\centering \caption{The variation with time of the effective
temperature around the time when peak temperature is achieved in
the TNR for the sequences on the 1.35 M$_\odot$ WD.  The time-scale
is identical to Figure 1 and shows how rapidly the
nuclear burning products are transported from the depths of the
hydrogen burning shell to the surface.  The different
evolutionary sequences are given on the plot.}
\end{figure}

The plots of the luminosity (L /L$_\odot$) over the first few hours of the 
outburst (Figure 5) demonstrate that, if we could observe a CN sufficiently early in the outburst,
then it should be super-Eddington. The initial spike, at a time of
about 100 s, is caused by a slowing of the expansion as the energy release
from the $\beta^+$-decays decreases.  After this time, expansion
and cooling of the outer layers causes the opacity to increase and
radiation pressure then accelerates the layers outward.  The
continuous flow of heat from the interior, combined with the
increase in opacity, causes another increase in luminosity until
the peak is reached. After this time, the layers continue expanding and
cooling until a radius of about $10^{12}$ cm is reached.  The effective
temperature has decreased to a value where most of the radiation is
being emitted in the optical and visual maximum occurs.

As reported in \citet{starrfieldpep09}, 
including only the sequences computed with the $\it pep$ reaction, if we
examine the abundance predictions for the 1.25 M$_\odot$ sequences,
we see that the differences caused by improving the reaction rate
library are small except for a few nuclides.  For example, the abundance of $^{12}$C
varies by about a factor of two, $^{14}$N by less than a factor of
two, and $^{16}$O by about a factor of 1.5.  The low mass odd-A
isotopes ($^{13}$C, $^{15}$N, and $^{17}$O) all vary by about a
factor of two.  However, both $^{12}$C and $^{13}$C are depleted
in the latest sequence as is $^{15}$N, while $^{17}$O is enriched
when using the latest reaction rate library.  The more massive
nuclides ejected in the 1.25 M$_\odot$ simulation, $^{22}$Na, $^{26}$Al, and
$^{27}$Al, are depleted, while
$^{32}$S is enhanced in the simulation done with the most modern library.  Tabulations of the ejecta
abundances are given in \citet{starrfieldpep09}.  Some of these abundances are
in good agreement with those measured in the Anomalous Interplanetary Particles as 
reported in \citet{pepin_2011_aa}.

\begin{figure}
\includegraphics[angle=90,scale=0.33]{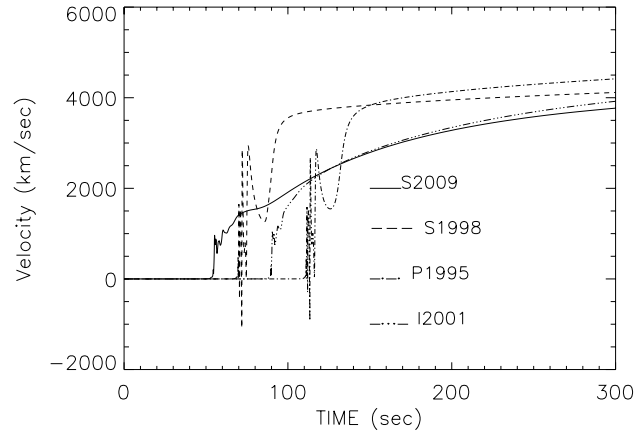}
\centering \caption{The variation with time, over the first 300
s of the outburst, for the velocity of the surface zone using
the four different reaction libraries.  We have offset the time on the \citet{politano_1995_aa}
sequence to make the curves more visible.}
\end{figure}
   
The effects of changing the nuclear reaction rate library are more
apparent for the sequences accreting onto 1.35 M$_\odot$ WDs.  Both $^{12}$C and
$^{13}$C drop in abundance using the latest library while $^{14}$N,
$^{16}$O and $^{17}$O increase in abundance.  In addition, there
is hardly any difference in the abundances as a function of WD
mass, except for $^{14}$N, which is lower by about a factor of 3 at
1.35M$_\odot$.  We also find that the abundance of $^{22}$Na
is lowest in the calculations done with the latest library.  However, it is still a factor of about 5
more abundant at 1.35 M$_\odot$ than at 1.25 M$_\odot$.  
The abundances of both $^{26}$Al and $^{27}$Al increase slightly from
1.25M$_\odot$ to 1.35 M$_\odot$.  A similar result is seen in \citet{jose_1998_aa}
although their ejecta abundances are smaller than ours because of
the difference in initial abundances and peak temperatures.  A detailed study of the formation of both $^{26}$Al and $^{27}$Al can be found in \citet{jose_1999_aa} who consider the abundances, reactions, and reaction rates that
take part in forming these nuclei.  In \citet{starrfield_1998_aa, starrfield_2000_aa} we reported that the
abundance of $^{26}$Al declined as the WD mass increased.  The simulations
presented  in those papers used older reaction rate libraries and did not
include the {\it pep} reaction so the temperature and isotopic evolution were
different.  Finally,
the abundance of $^{32}$S is largest using the 2005
library at 1.35 M$_\odot$. In fact, it reaches 4\% by mass of the ejected
material.  This result may, in part, explain the large sulfur abundance found
for V838 Her \citep{vanlandingham_1996_aa}.

\begin{figure}
\includegraphics[angle=90,scale=0.33]{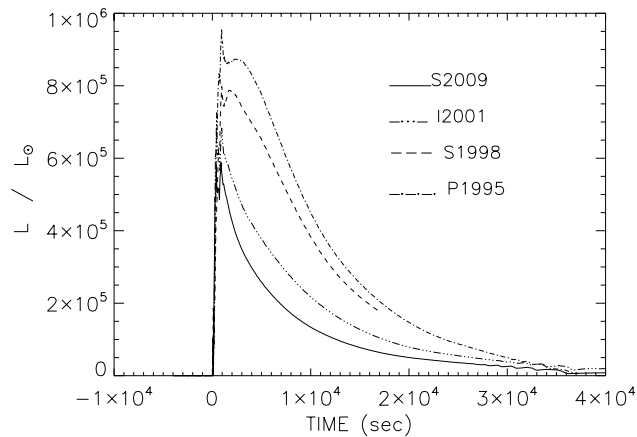}
\centering \caption{The variation in time, over the first 11 hours
of the outburst, of the surface luminosity using the four
different reaction libraries. The label which identifies each
different sequence is given in the legend. Note that as the nuclear
physics input has improved, the peak luminosity and the luminosity at
later times has decreased.}
\end{figure}

In all four
sequences the ejected total oxygen abundance exceeds the carbon abundance as found in our
earlier studies.  This result continues to be puzzling in light of
the production of carbon rich dust in CN ejecta. As described in
\citet{gehrz_1998_aa} and \citet{jose_2004_aa} infrared studies
suggest that C $>$ O is required for the formation of SiC and
amorphous carbon grains since the CO molecule forms in the
ejecta and is very stable.  This implies that it is only the left over
carbon that is available for grain formation.  On the other hand,
if O $>$ C, then the leftover oxygen goes to form oxides and 
silicates.  Yet, some of the deep dust forming novae form all
types of dust since C, SiC, and hydrocarbons were identified early in the
observed outburst, and O-rich silicate grains later in the outburst, for both
QV Vul and V705 Cas \citep{gehrz_1998_aa, woodward_2011_aa}, which suggests that distinct regions
with O $>$ C and O $<$ C can occur in the ejecta of the same
CN \citep{rawlings_1995_aa, rawlings_2002_aa}.

This question
has also been investigated by both \citet{shore_2004_aa} and
\citet{jose_2004_aa}.   \citet{shore_2004_aa} investigated
the effects of the UV radiation field on grain formation and
suggested ``a possible formation mechanism for large grains: ionization-mediated kinetic 
agglomeration of atoms onto molecules and small grains
through induced dipole interactions."
\citet{jose_2004_aa} reported on the results from 
1-D hydrodynamic simulations of CN outbursts and then used their
isotopic results in calculating thermodynamic equilibrium 
condensation sequences for the ejecta.  They
studied both CO novae and ONe novae and, in some detail, 
the influence of other elements such as Al, Ca, Mg, and Si.  
Interestingly, they found that SiC grains are likely to condense in
ONe novae.  However, their results for CO novae show that 
SiC grains do not form in the ejecta and therefore these novae
do not contribute to pre-solar grains.  They also regard the 
formation of carbon dust in CO novae as still a puzzle although
they state that it is possible to form carbon rich grains even in
an environment with O $>$ C.  Further work in this area is 
warranted.

Since \citet{starrfieldpep09} was published, further improvements to calculate nuclear reaction rates useful for
astrophysics have been
done by Iliadis and collaborators, as summarized in the
paper introducing STARLIB \citep{sallaska_2013_aa}.  
STARLIB is a tabular, stellar reaction rate library that includes neutrons, 
protons, $\alpha$-particles, $\gamma$-rays, and nuclides ranging from Z = 1 to 83.  
All available experimental nuclear physics information is used to compute the rates.  The structure of STARLIB rests on a Monte Carlo method to quantitatively define reaction rate uncertainties 
\citep{longland_2010_aa, iliadis_2010_aa, iliadis_2010_ab, Iliadis_2010_ac}.  The method uses experimentally determined nuclear physics quantities (resonance strengths and energies, S-factors, partial widths, etc.) as inputs to a Monte Carlo algorithm.   Full details of the Monte Carlo method can be found in \citet[][and references therein]{sallaska_2013_aa} and \citet{iliadis_2015_aa}.  Here we only highlight the basics of the procedure.  All the measured nuclear physics (input) properties entering into the reaction rate calculation are randomly sampled according to their individual probability density functions. The sampling is repeated many times and thus provides the Monte Carlo reaction rate (output) probability density. Finally, the associated cumulative distribution is determined and is used to define reaction rates and their uncertainties with a precise statistical meaning (i.e., a quantifiable coverage probability).  For example, for a coverage probability of 68\%, the low, recommended, and high Monte Carlo rates can be defined as the 16th, 50th, and 84th percentiles, respectively, of the cumulative reaction rate distribution.  
 
STARLIB contains experimental Monte Carlo rates for 62 charged-particle nuclear reactions on A = 14 to 40 target nuclei \citep{iliadis_2010_aa}. In \citet{sallaska_2013_aa} seven updated Monte Carlo rates were reported, plus one entirely new rate.  What is interesting for CN  simulations is that experimental Monte Carlo rates are available for almost all reactions participating in the TNR. The situation is drastically different from other stellar explosions, such as supernovae, where the rates of most reactions are based on nuclear theoretical models such as Hauser-Feshback.

\begin{figure}
\includegraphics[scale=0.33]{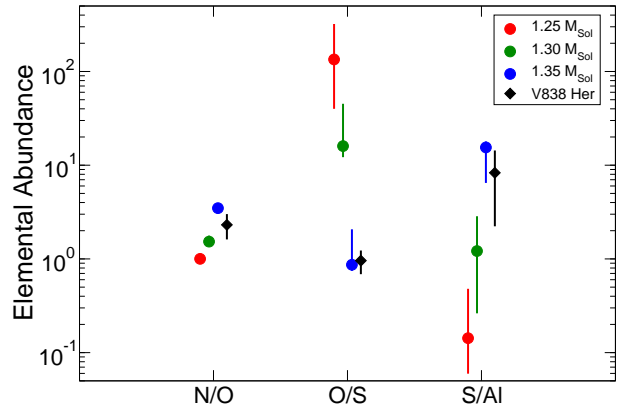}
\centering \caption{Abundance ratios of the nuclear thermometers N/O, O/S, and S/Al, for V838 Her.
The circles in color are from model simulations involving different WD masses,
and the purple diamonds show the observed values. The results show that V838 Her
involved a WD with a mass close to 1.35 M$_\odot$.}
\end{figure}

The STARLIB reaction rates have now been used in two papers relevant to studies of the Classical Nova Outburst.
\citet{downen_2013_aa} generated a series of new classical nova simulations using SHIVA \citep{jose_1998_aa}
with reaction rates adopted from STARLIB.  Evolutionary sequences were generated for WD masses
ranging from 1.15 M$_\odot$ to 1.35 M$_\odot$, and the parameters describing both the initial models and evolutionary results were given in the paper.  They adopted the temperature-density-time trajectories from the
hydrodynamic calculations and post-processed them with an extended nuclear reaction network. 
Their sequences reached peak temperatures ranging from
$2.28 \times 10^8$K to $3.13 \times 10^8$K, depending upon WD mass.   These temperatures are thought to be typical for simulations of TNRs involving ONe WDs, but are lower than the peak temperatures reached in the sequences reported in \citet{starrfieldpep09} and shown in Figure 1.  This difference in peak temperatures is caused by the lower initial $^{12}$C abundance used in \citet{starrfieldpep09}.

Carbon initiates CNO burning, because the $^{12}$C(p,$\gamma$) rate is considerably faster than
either the $^{14}$N(p,$\gamma$) rate or the $^{16}$O(p,$\gamma$) rate (higher Coulomb penetrability), so a larger initial carbon abundance causes the TNR to occur earlier with less material accreted and a lower electron degeneracy \citep[][Starrfield et al. 2016, in preparation]{jose_1998_aa}.  \citet{downen_2013_aa} assumed a mixing fraction of 50\% between accreted matter of Solar composition \citep{lodders_2009_aa} and WD core material prior to the outburst.  The abundances of core matter are taken from the evolution of a 10M$_\odot$ star from the main sequence to the end of core carbon burning \citep{ritossa_1996_aa} and are reported in Table 3 of \citet{downen_2013_aa}. 
We also note that \citet{downen_2013_aa} provided an updated table of observed abundances for ONe novae, since
the last detailed table of classical nova abundances was provided in \citet{gehrz_1998_aa}.

\citet{downen_2013_aa} used the post-processing results to compute a number of elemental abundance ratios that 
they could compare to the observed abundances of several well-studied novae.  As shown in Figure 6, they used the N/O and O/S ratios from their post-processing to predict that the peak temperature in V838 Her was $\sim 3 \times 10^8$K and the WD mass in this system was $\sim 1.35$ M$_\odot$.  For V382 Vel, they predicted a peak temperature $\sim 2.3 \times 10^8$K and a WD mass of $\sim 1.2$ M$_\odot$.  However the results for the other novae that they studied (V693 Aql, LMC 1990\#1, V1065 Cen, and QU Vul) were less clear and no significant predictions could be made.  They concluded that the elemental ratios N/O, N/Al, O/Na, and Na/Al are robust in that they do not depend significantly on uncertain reaction rates.

\citet{kelly_2013_aa} continued these studies by varying the amount of mixing of accreted with core material from
values of 25\% to 75\%.  They used the same 4 WD masses as before but the highest peak temperature,
$3.44\times 10^8$K, was achieved in the simulation that assumed 75\% mixing of WD core material into the accreted envelope on a WD with a mass of 1.35 M$_\odot$.   Instead of varying reaction rates one-by-one, they employed a Monte Carlo reaction network method, where many reaction network samples were computed. For each calculation the rates of all reactions were randomly sampled, according to their probability densities listed in STARLIB.  This method is described in more detail in \citet{longland_2012_aa} and \citet{iliadis_2015_aa}. They again searched for those elemental abundance ratios that could be used to determine the amount of mixing of accreted with WD core matter.  As seen in Figure 7, using the ratios of Ne/H, Mg/H, and Al/H they found for V838 Her, V4160 Sgr, and V1974 Cyg that $\sim$ 25\% of outer WD core matter was mixed into the envelope prior to the TNR, contrary to the most common assumption of a 50\% mixing fraction. They find roughly the same percentage for LMC 1990\#1 and V693 CrA.  For V1065 Cen, a larger ratio of 50\% may apply.  In contrast, the results for V382 Vel and V1974 Cyg are in poor agreement with any of the predicted elemental ratios. They also found poor agreement between observed and predicted abundances for other ONe novae, e.g., V838 Her and QU Vul.  Their results suggest that new studies of TNRs for ONe novae should be done with smaller values for the amount of core material mixed into accreted material.  This result is supported by the multi-D models of pre-outburst mixing \citep{casanova_2010_ab, casanova_2011_ab} which mix 30\% of core material into the envelope.

\begin{figure}
\includegraphics[scale=0.33]{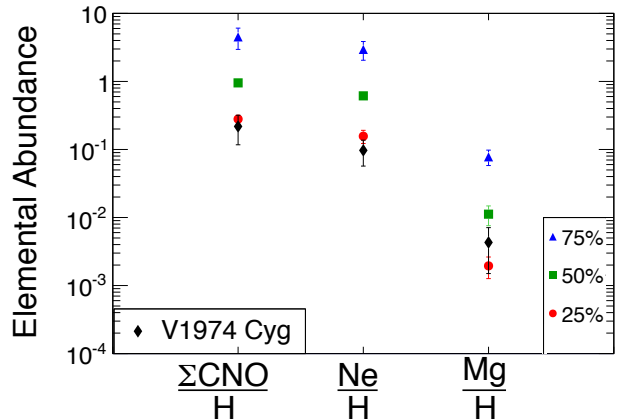}
\centering \caption{Abundance ratios of the mixing meters CNO/H, Ne/H, and Mg/H for V1974 Cyg. The circles in color correspond to model simulations for different degrees of pre-enrichment (25\%, 50\%, 75\%). The black symbols
show the observed values. The results indicate that 25\% of ONe WD core matter was mixed into the envelope
for this outburst.}
\end{figure}

\section{Multidimensional Studies of the Thermonuclear Runaway}

Despite great efforts over the past several decades, two interconnected
problems continue to plague our understanding of 
thermonuclear processes in the CN outburst.  They are (1) how does the convective
region grow and develope in
response to the TNR, and (2) how and when are the WD core nuclei
mixed into the accreted matter?  As already shown in the one
dimensional hydrodynamic calculations, the transport of heat and
$\beta^+$-decay nuclei to the surface by convection, as the TNR
rises to its peak, is extremely rapid and may influence a number of
observable features of the CN outburst that can be used both to
guide and constrain new simulations.  The first is the early evolution 
of the visual light curves of fast CNe on
which their use as ``standard candles'' is based. During this
phase the bolometric luminosity of a nova can remain more than an
order of magnitude above the Eddington luminosity for several days
(observed for LMC 1991 by Schwarz et al. 2001).  The
second is the composition of matter ejected by a nova as a function of time. It
is possible that material ejected early in the outburst may not
have the same composition (isotopic or elemental) as material
ejected later. Both of these features depend on the amount, timing, and
composition of the material dredged up from the underlying CO or
ONe WD core.

It has now become possible to treat convection at or near the peak of the TNR in
both two and three dimensions.  This is possible for CN studies
since the relevant timescales are all on the order of seconds. For
example, the dynamical timescale $\tau_{hyd}$, at a density of
$\sim$10$^4$ g cm$^{-3}$ is of the order of seconds. The nuclear burning
timescale decreases from years to seconds, once the temperature
rises above 10$^8$ K, until constrained by the $\beta^+$-decay
lifetimes. Finally, the convective  turn-over timescale is of
order of seconds near the peak of the runaway \citep{starrfield_1998_aa}.

\citet{fryxell_1982_aa}  first
discussed the importance of multidimensional effects for TNRs that
occurred in thin stellar shells. For CNe, they assumed initiation
at a point and  calculated the lateral burning velocity of the
deflagration front that spread the burning along the surface.
Subsequently, \citet{shankar_1992_aa} and \citet{shankar_1994_aa} 
carried out two dimensional hydrodynamic calculations of this
problem. They restricted their survey to strong, instantaneous,
temperature fluctuations that developed on a dynamical time scale.
However, they found that the initially intense burning at a point
extinguished on a short timescale,  as the perturbed region
rapidly rose, expanded, and cooled.

\citet{glasner_1997_aa} explored the consequences of thermonuclear
ignition and explosive hydrogen burning in CNe with a two
dimensional, fully implicit hydrodynamic code. They followed the
evolution of a convectively unstable hydrogen-rich envelope
accreted onto a CO WD at a time close to the peak of the TNR and
found a flow pattern that effectively dredged up sufficient
material from the core to explain the observed levels of heavy
element enrichment in CNe ejecta 
\citep[$\sim$ 30\% to 40\% by mass:][]{gehrz_1998_aa, downen_2012_aa}.  
The redistribution of nuclear energy generation over the
envelope, caused by the outward transport of short lived
$\beta^+$-decay nuclei, was also found to play a significant role
in the outburst. In a complementary study, \citet{kercek_1998_aa}
examined  the early stages of the evolution, using the same
initial model as \citet{glasner_1997_aa} but with an explicit, Eulerian, hydrodynamic
code.  While their simulations confirmed
the finding of \citet{glasner_1997_aa},  mixing  was not as strong and
occurred over a longer timescale.

\citet{kercek_1999_aa} then performed two and three dimensional
studies, using the same input model and physics as before, but
with improved resolution.  Their results displayed less mixing
with core material and a completely different flow structure, which
cast doubts on this mixing mechanism. 
\citet{glasner_2005_aa}, however, analyzed the effects of the
surface boundary condition on the multidimensional calculations
and concluded that Lagrangian simulations, where the 
mass of the envelope matter is conserved and which allowed the 
outer boundary to expand, resulted in
explosions.  In contrast, Eulerian methods, where material is
allowed to flow off the numerical grid, did not result in
explosions.  However, \citet{casanova_2010_ab,
casanova_2011_aa, casanova_2011_ab} used FLASH
\citep{fryxell_2000_aa}, an Eulerian, explicit code, to study mixing in 3
dimensions and did find explosions in which the ejecta were enriched by $\sim$30\% in
core matter.  \citet{glasner_2012_aa} performed
2-dimensional calculations on WDs with different
core compositions and again found mixing via convective dredge-up.

Earlier, however, \citet{rosner_2001_aa} re-examined shear mixing 
\citep{kippenhahn_1978_aa, sparks_1987_aa, kuttersparks_1987_aa}  
during the accretion phase and suggested, based on
semi-analytical and timescale arguments plus two dimensional
calculations, that this mechanism could also be responsible for
significant mixing.  \citet{alexakis_2004_aa} studied the development of
shear mixing on a 1.0M$_\odot$ WD in two dimensions. Their initial
model consisted of a completely convective layer moving at a large
velocity tangential to the surface of the WD.  However, they did not include
nuclear burning. They found core material was mixed into the
envelope, but they failed to address how such a thick layer could
have formed on the surface of the WD since the accreting material
must have mixed far earlier in the evolution. In addition, since
they mixed CO material into a H-rich layer with a peak temperature
of $10^8$K, they would have obtained an explosion if nuclear burning
had been included.

\citet{walder_2008_aa} using their own code \citep{walder_2000_aa}, which is ``a 
parallel, block-structured, adaptive mesh refinement (AMR) hydrodynamical code using Cartesian meshes and multidimensional high-resolution finite-volume integration'', followed 
the 3-D  evolution of both the accretion and explosion phase of RS Oph.  This system consists
of a WD exploding inside the outer layers of a red giant and their work provided a detailed look
at the external shock moving through the red giant atmosphere.  Based on their simulations,
they concluded that the WD in RS Oph was increasing in mass and evolving toward a SN Ia
explosion. 

As already mentioned above, the most recent multidimensional studies of the CN outburst were
performed by \citet{casanova_2010_ab, casanova_2011_aa, casanova_2011_ab},
who re-investigated
the 2-D simulations originally reported by \citet{kercek_1999_aa}
and \citet{glasner_1997_aa} and continued with 3-D simulations.  
They used the FLASH code \citep{fryxell_2000_aa}
and showed that an Eulerian formulation, with sufficient resolution and the proper boundary
conditions, produced sufficient mixing to agree with the observations. \citet{casanova_2011_ab}
reported that the mixing occurred from the action of the Kelvin-Helmholtz instability
driving mixing across the accreted material core boundary.  Therefore, the 3-D studies reported in \citet{casanova_2011_ab} showed 
conclusively that the only way to treat convective mixing is in 3-D and they described in
detail why 2-D simulations of convection are ``unrealistic''\citep[see also][]{arnett_2014_aa,arnett_2015_aa}.
A recent summary of their findings is given in \citet{jose_2014_aa}.

Given these calculations, the general
inferences that can be drawn from the existing multidimensional
calculations are that: (1) the amount of mixing
occurring prior to the onset of convection in the TNR is negligible; (2) the amount of
convective mixing occurring during the early stages of the TNR is
a sensitive function of the degree of degeneracy; (3)
Kelvin-Helmholtz driven convective mixing dredges-up sufficient CO- or ONe-rich
matter from the underlying WD core to produce the observed enrichments of nova ejecta \citep{jose_2014_aa};
and (4) Since the heavy element enrichment of the envelope via 
dredged up material, does not occur until after convection has been initiated in the
nuclear burning regime, late in the evolution of the TNR,
the envelope composition during accretion is that of the material
being transferred by the secondary.  This keeps core material out
of the accreted layers until the peak of the TNR, the
opacity stays low, and the amount of accreted material is increased
\citep{starrfield_1998_aa}.

\section{Nucleosynthesis during the TNR and the Mass of the Ejecta}

The measured abundances for CN ejecta confirm the levels of
enrichment required by the theoretical studies to reproduce the
dynamic features of CNe outbursts and, in addition, establish
that both CO and ONe WDs occur in cataclysmic variable binary
systems \citep{gehrz_1998_aa, starrfield_1998_aa, starrfield_2008_CN}. 
Further, the significant enhancements of heavy
elements in CN ejecta, taken together with the observational
determinations of the masses of their ejecta, confirm that CNe
contribute significantly to the Galactic abundances of some CNO isotopes.
Finally, possible signatures of nova processing have already been identified in
pre-solar grains found in meteorites \citep{amari_2001_aa, jose_2004_aa} and in the Anomalous
Interplanetary Particles \citep{pepin_2011_aa}.

The extensive database of atmospheric and nebular elemental
abundances for CNe ejecta \citep{gehrz_1998_aa, starrfield_1998_aa,downen_2013_aa}
constitutes a
powerful tool both for constraining the modeling of their
outbursts and for determining their contributions to Galactic
chemical evolution. The degree to which elements such as silicon,
sulfur, and argon are enriched, according to previous
nucleosynthesis studies, is a sensitive function of the
temperature history of the burning shell as are the abundances of $^{22}$Na and $^{26}$Al
 \citep{jose_1997_aa, jose_1998_aa, jose_1999_aa, jose_2001_ab, starrfield_2001_aa, iliadis_2002_aa, hix_2003_aa, paretekoon_2003_aa, yaron_2005_aa, starrfieldpep09}.  Finally, the
abundance of $^7$Be in CN ejecta is sensitive to the rate at which
it is transported to the surface regions prior to its decay to
$^7$Li. Until recently the CN contributions to the abundance  of $^7$Li
in the Galaxy \citep{starrfield_1978_aa, hernanz_1996_aa}, and expectations for the detection of $\gamma$-rays
from $^7$Be decay in CN ejecta, remained open questions \citep{romano_1999_aa, romano_2003_ab,
romano_2003_aa}.  However, the advent of high dispersion spectroscopy of CN early in the outburst
has resulted in the discovery of $^7$Be in V339 Del \citep{tajitsu_2015_aa} and $^7$Li in V1369 Cen \citep{izzo_2015_aa}.  The abundance of $^7$Li as measured by \citet{izzo_2015_aa} results in a lithium mass
of a few times $10^{-10}$M$_\odot$ from a single nova and they suggest ``that this amount solves the
origin of the overabundance of lithium observed in young stellar populations.''

Since both CO and ONe WDs are found in CN systems, it is crucial
to calculate evolutionary sequences for consistent choices of WD
mass, envelope mass, thermal structure, and composition (CO or
ONe) that can be compared directly to observed CN
systems. While we have evolved one dimensional sequences designed
to fit the observed properties of the ONe nova V1974 Cyg \citep{starrfield_2000_aa},
these simulations predicted sufficient $^{22}$Na production 
that its $\gamma$-ray emission should have been
detected by the Imaging Compton Telescope (COMPTEL), but it was not detected \citep{shrader_1994_aa, iyudin_1995_aa, leising_1997_ab,
leising_1997_aa}.  Another discrepancy was that
comparison of the abundance predictions with observations
suggested that the \citet{starrfield_1998_aa} simulations were over-producing nuclides in
the mass region past magnesium. One source of these discrepancies
appears to be the use of the post carbon burning abundances of
\citet{arnett_1969_ab} for their WD core abundances.  In contrast,
the study of carbon burning nucleosynthesis by \citet{ritossa_1996_aa} 
predicted lower abundances for Mg and Si.  If this
composition is implemented, then a lower level of $^{22}$Na
production is obtained and there is no contradiction
with the lack of detection of  $\gamma$-rays from $^{22}$Na
decays \citep{jose_1998_aa, starrfield_2000_aa, starrfieldpep09}. 
However, while nuclear decay $\gamma$ -rays have not been seen, the \textit{Fermi}/Large Area Telescope (LAT) has discovered that four CNe, are $\gamma$-ray sources at E $>$ 100 MeV (V959 Mon 2012, V1324 Sco 2012, V339 Del 2013, V1369 Cen 2013) in the earliest stages of their outbursts \citep{ackermann_2014_aa}.

Another long-standing problem is the discrepancy between
observations and predictions of the amount of mass ejected in the
outburst \citep{warner_1995_aa, starrfield_1998_aa, starrfield_2000_aa}. IR and radio analyses, combined
with optical and UV studies of nebular emission lines, provide
estimates of the ejected mass \citep{warner_1995_aa, gehrz_1998_aa}. 
In contrast to the observationally determined masses,
numerical simulations of TNRs on both CO and ONe WDs predict
ejecta masses that can be smaller by up to a factor $\sim$10
\citep{prialnik_1995_aa,  jose_1999_aa, starrfield_1998_aa, starrfield_2000_aa, yaron_2005_aa, starrfieldpep09}.

There are two reasons that the cause of this ejecta mass
discrepancy must be determined. First, a solution should provide
an improved understanding of the development of the CN outburst; and
second, most estimates of the contributions of CNe to Galactic
chemical evolution use ejecta masses determined from the
theoretical predictions.  If the masses inferred from observations
are used in the chemical evolution studies, then CNe become even
more important for production of the odd isotopes of the light
elements in the Galaxy (particularly $^{13}$C, $^{15}$N, and
$^{17}$O) than currently believed.

\section{Can Classical Novae be Progenitors of Type Ia Supernovae?}

Supernovae of Type Ia (SNe Ia) are those supernovae in which
neither hydrogen nor helium is seen in any of the spectra obtained during
the outburst.  However, the large sample size of recent SN surveys have revealed
that there are occasional events that are otherwise SN Ia but, in fact, they do show
small amounts of H in their spectra.  SN Ia have light curves that can be
calibrated \citep{phillips_1993_aa}, making them excellent standardizable distance indicators
to {\it z} $>$ 1\citep{filippenko_1997_aa, howell_2010_ab}. Thus, they have become
extremely important tools to determine the
structure and evolution of the Universe
\cite[][and references therein]{leibundgut_2000_aa,leibundgut_2001_aa}. 
SN Ia are also important because they
contribute a major fraction of the iron group elements to the
Galaxy.  In the past few years, a tremendous effort has gone into
studies of their observed
properties \citep[cf.,][]{hillebrandt_2003_aa, howell_2010_ab, maoz_2014_aa, ruiz_2014_aa}. Nevertheless,
the progenitor(s) of SN Ia explosions are, as yet, unknown. 

\citet{whelan_1973_aa} 
proposed that the explosion involved a CO WD
which accreted material from a binary companion until its mass
approached the Chandrasekhar Limit and a carbon
deflagration/detonation occurred \citep{nomoto_1984_aa, branch_1995_aa, 
hillebrandt_2000_aa}.  Typical CN systems can be excluded as SN progenitors because the WD is
thought to be {\it decreasing} in mass as a result of repeated nova explosions and
cannot be growing toward the Chandrasekhar Limit
 \citep{macdonald_1984_aa, gehrz_1998_aa, starrfield_2000_aa}.  
In addition, the absence of hydrogen and helium in the
spectra of a SN Ia rules out most other Cataclysmic Variable (CV) systems since the WDs
are accreting hydrogen- and helium-rich material. If the WD were
to explode, then the accreted envelope would be carried along with
the supernova ejecta and be seen in the spectrum \citep{marietta_2000_aa,  starrfield_2003_aa}.

Nevertheless, one suggestion for the progenitors of SN Ia explosions 
is the transfer of matter from a non-degenerate secondary onto a
WD in a close binary system.  If a sufficient amount of the 
accreted material remains on the WD, during the accretion process
and its mass can gradually grow close to the Chandrasekhar Limit,
then the explosion should 
resemble a SN Ia.  This hypothesis is referred to as
the single degenerate scenario (SD).  It is one of the two major
suggestions for possible progenitors of SN Ia explosions, the other
being the double degenerate (DD) scenario.  In the SD scenario, as the
WD in a close binary system approaches the Chandrasekhar Limit, it first convectively
``simmers'' in the core and then the explosion occurs.  In contrast, the
double degenerate scenario (DD) requires the merger or collision of
two WDs to produce the observed explosion.  While there are now major
efforts to better understand the DD scenario, the SD scenario is
capable of explaining most of the observed properties of the SN Ia
explosion via the delayed detonation hypothesis 
\citep[and references therein]{khokhlov_1991_aa, kasen_2009_aa, woosley_kasen_11_a,
  howell_2009_ab}.  Reviews of the various proposals for SN Ia
progenitors \citep{branch_1995_aa}, and the
implications of their explosions can be found in
\citet{hillebrandt_2000_aa},
\citet{leibundgut_2000_aa,leibundgut_2001_aa}, \citet{nomoto_2003_aa},
and \citet{howell_2010_ab}. Recent reviews of the observations can be found in \citet{maoz_2014_aa}
and \citet{ruiz_2014_aa}.

New evidence in favor of the SD scenario comes
from observations of SN 2011fe in M101.  They imply that 
the exploding star was likely a CO WD
\citep{pnugent11} with a companion that was probably on or near the main sequence
\citep{weidongli_2011_aa,bloom_2012_aa}.  However, EVLA
\citep{chomiuk_2012_aa} and optical \citep{bloom_2012_aa} observations
may have ruled out many types of CVs.  In addition,
HST studies of the spatial region from which SN 2011fe exploded, suggest that
the progenitor had a luminosity less than 
$\sim10^{34}$ erg s$^{-1}$ 
\citep{graur_2014_aa}, and \citet{lundqvist_2015_aa} find no evidence for
a remnant companion in late time observations of SN 2011fe and SN 2014J.  While
this rules out typical Supersoft X-ray sources \citep{kahabka_1997_aa},
recent studies suggest that a CV progenitor could be fainter than that value
\citep{newsham_2014_aa, starrfield_2012_basi, starrfield_2014_aa}.

Moreover,
\citet{dilday_2012_aa} claim that PTF 11kx was a SN Ia that exploded
in a Symbiotic Nova system.  Finally, we note that the ``zoo'' of SNe Ia types is
increasing as surveys find more and more members \citep[e.g.,][]{white_2015_aa}.  
The most recent results \citep{cao_2015_aa,
olling_2015_aa}, {\it in the same issue of Nature},  both favor and disfavor the
SD scenario.  Therefore, since the existence
of ``Super-Chandra'' Ia's suggests that DD mergers are required for
these extreme explosions,  these studies taken together suggest that
there are multiple channels that can produce SN Ia explosions 
including the SD channel.

Further support for the SD channel, comes from the observations of V445
Pup (Nova 2000).  There were no signs of hydrogen in the
spectrum at any time during the outburst, especially just after discovery, but there were strong lines
of carbon, helium, and other elements in the optically thick
spectra \citep[]{wagner_2001_aa, wagner_2001_ab, henden_2001_aa,
lyke_2001_aa, woudt_2005_aa,
 woudt_2009_aa}.  Unfortunately, no one has done an abundance
 analysis of the spectra,  obtained early in the outburst, to determine an upper limit to the amount of hydrogen
 that could be hidden.  Nevertheless,
 it is probably extremely small.  Because it was extremely luminous before the
outburst, the secondary is thought to be a hydrogen deficient carbon
star \citep[]{woudt_2009_aa}.  Since one of the defining
characteristics of a SN Ia explosion is the absence of hydrogen or
helium in the spectrum at any time during the outburst or decline, the
existence of V445 Pup implies that mass transferring binaries exist in
which hydrogen is absent at the time of the explosion and most of the
helium is converted to carbon during the nova phase of evolution.  The latest
spectra show that this system is still in outburst and, therefore, it
has not been possible to study the underlying system\citep{tomov_2015_aa}.

In order to simulate the properties of the Super Soft X-ray Sources and determine if they
could be SN Ia progenitors, \citet{starrfield_2004_aa} 
used NOVA to study
accretion onto hot, luminous WDs and found that hydrogen
burns to helium (and helium to carbon and oxygen) in the surface
layers for a broad range of mass accretion rates.   They reported that accretion,
(from $1.6 \times 10^{-9}$ M$_\odot$yr$^{-1}<$ \.M $< 8 \times
10^{-7}$ M$_\odot$yr$^{-1}$), onto hot ($2.3 \times 10^{5}$ K),
luminous (30 L$_\odot$), massive (1.25 M$_\odot$, 1.35 M$_\odot$) CO
WDs could burn matter at rates both
lower and higher than the single value assumed in the canonical Steady
Burning scenario.  In addition, because of the energy
release from hydrogen burning near the surface, \citet{starrfield_2004_aa} found that the
helium layer remained hot, and helium steadily burned to carbon,
oxygen, and more massive nuclei without experiencing a TNR.
No mass was ejected and the WD grew in mass toward the Chandrasekhar
Limit. Some sequences were evolved for more than $10^6$yr. 
Since most of the hydrogen and helium accreted from the
secondary burned to carbon and oxygen, there would be almost no
hydrogen or helium present in the ejecta (and spectrum) if the WDs exploded as a
SN Ia.  In addition, the luminosities
and effective temperatures of their evolutionary sequences fit the
observations of the Super Soft X-ray Sources such as CAL 83 and CAL 87. 

In addition, because the multidimensional studies imply that mixing of core with
accreted material does not occur until after convection is occurring just prior to the
peak of the TNR \citep[][and references therein]{jose_2014_aa},
we have also investigated the accretion of Solar material onto WDs of
various masses and mass accretion rates \citep{starrfield_2012_basi, starrfield_2012_aa,
newsham_2014_aa}.  We found that in all cases the simulation evolves to a TNR but only a
small amount of accreted material is ejected and the WDs are growing in mass.  These
studies used both NOVA and MESA \citep{paxton_2011_aa, paxton_2013_aa, paxton_2015_aa} and show that the evolutionary sequences exhibit the \citet{schwarzschild_1965_aa} thin shell instability, which implies that steady burning does not occur.  An expanded study of the stability of thin shells can be found in \citet[][]{yoon_2004_aa}, who investigated
the accretion of hydrogen-rich material onto WDs.  Using their results, we find that our sequences begin in their stable region, but with continued accretion, evolve into instability.   
The high mass WDs do eject a small fraction of the accreted material (a maximum of $\sim4\%$ for the 1.25M$_\odot$  sequences, but only $\sim0.1\%$ for the 0.7M$_\odot$ sequences).  
We identify these systems with those CVs (dwarf, recurrent, symbiotic
novae) that show no core material either on the surface of the WD or
in their ejecta.  Our results could explain the findings of
\citet{Zorotovic_2011_aa}, who report that the WDs in CVs are growing in
mass.  In addition, the best studied dwarf novae have WD masses larger
than the canonical value of $\sim$0.6M$_\odot$.  These are 
U Gem \citep[1.2M$_\odot$:][]{Echevarria_2007_aa}, 
SS Cyg \citep[0.8M$_\odot$:][]{Sion_2010_SSCyg_aa}, 
IP Peg \citep[1.16M$_\odot$:][]{copperwheat_2010_aa}, and 
Z Cam \citep[0.99M$_\odot$:][]{Shafter_1983_aa}.  Therefore, it seems possible
that some Dwarf Novae could be SN Ia progenitors if there is some means to
prevent convection from mixing accreted with WD core material.  For example, a thick
helium layer from previous outbursts could act as a barrier to this mixing. Further
work is warranted.

\section{Summary and Discussion}

We have reviewed our current understanding of the
thermonuclear processing that occurs during the evolution of the
CN outburst.  A TNR in the accreted hydrogen-rich layers on the
{\it low} luminosity WDs in Cataclysmic Variable binary systems is
the outburst mechanism for Classical, Recurrent, and Symbiotic
Novae. The interaction between the hydrodynamic evolution and
nuclear physics lies at the basis of our understanding of how the
TNR is initiated, evolves, and grows to the peak of the explosion.
The {\it observed} high levels of enrichment of CN ejecta in
elements ranging from carbon to sulfur confirm that there is
dredge-up of matter from the core of the WD and enable CNe to
contribute to the chemical enrichment of the interstellar medium.
Therefore, studies of CN are leading to an improved understanding
of Galactic nucleosynthesis, the sources of pre-solar grains, the
extragalactic distance scale, and the nature of the progenitors of
SN Ia.

It is now recognized that the characteristics of the CN explosion
depend on the complex interaction between nuclear physics (the
$\beta^+$-limited CNO cycles) and convection during both the early and final
stages of the TNR. The light curves, the peak luminosities (which
can exceed the Eddington luminosity), the levels of envelope
enrichment, and the composition of CN ejecta are all strongly
dependent upon the extent and timescale of convective mixing
during the explosion. The characteristics of the outburst depend
upon the WD mass, WD luminosity, mass accretion
rate, the chemical composition of both the accreting and
WD core material, the evolutionary history of the WD,
and when and how the accreted layers are mixed with the
WD core. The importance of nuclear physics to our
understanding of the progress of the outburst can be seen when we
compare a series of evolutionary sequences in which the only
change has been the underlying nuclear reaction rate library.
In order to make meaningful comparisons of theory
with observations, we need to use the best nuclear physics and
opacities that are available. 

We have also highlighted a number of
problems in our understanding of the outburst. 
Prominent among them are the timing and quantity of mixing of WD
material into the accreted layers and the discrepancy between the 
theoretical predictions of the amount of material ejected during
the outburst and the observations.  These problems have
a number of important implications and must be solved before we
can claim a better understanding of the outburst.

We are extremely grateful to the referee for a detailed and careful
reading of the original manuscript which has improved the
presentation.
We are also grateful to a number of collaborators who over the years
have helped us to better understand the nova outburst.  We have
benefitted from discussions with A. Beardmore, M. Bode, A. Champagne, M. Darnley, J. Drake, A. Evans, R. D.
Gehrz, P. H. Hauschildt, M. Hernanz, R. Hounsell, J. Jos\'e, S. Kafka, J. Krautter, T. Liimits, J.-U.
Ness, J. Osborne, K. Page, D. Prialnik, F. Sarina, G. Schwarz, H. Schatz, A. Shafter, G. Shaviv, S.N. Shore, E.
M. Sion, W. M. Sparks, P. Szkody, J. Truran, K. Vanlandingham,
R. M. Wagner, M. Wiescher, P. Woudt, and C. E. Woodward. 
SS acknowledges partial support from
NASA and NSF grants to ASU.  WRH acknowledges partial support from
DoE and NSF.  Oak Ridge National Laboratory is managed by
UT-Battelle, LLC, for the U.S. Department of Energy under contract
DE-AC05-00OR22725. CI acknowledges partial support by 
the U.S. Department of Energy under Contract no. DE-FG02-97ER41041.

\bibliography{references_iliadis,starrfield_master}

\end{document}